\documentclass{ws-jai}
\usepackage[flushleft]{threeparttable}
\usepackage{natbib}

\begin{document}

\catchline{}{}{}{}{} 

\markboth{McCoy et al.}{A primer for telemetry interfacing in accordance with NASA standards using low cost FPGAs}

\title{A primer for telemetry interfacing in accordance with NASA standards using\\ low cost FPGAs}

\author{Jake McCoy$^{\dagger \S}$, Ted Schultz$^\dagger$, James Tutt$^\dagger$, Thomas Rogers$^\ddagger$, Drew Miles$^\dagger$,  Randall McEntaffer$^\dagger$}

\address{
$^\dagger$University of Iowa, Van Allen Hall, Iowa City, IA, USA 52242, $^\S$jake-mccoy@uiowa.edu \\
$^\ddagger$University of Colorado, Center for Astrophysics and Space Astronomy, Boulder, CO, USA 80303
}

\maketitle

\corres{$^\S$Corresponding author.}

\begin{history}
\received{(to be inserted by publisher)};
\revised{(to be inserted by publisher)};
\accepted{(to be inserted by publisher)};
\end{history}

\begin{abstract}
Photon counting detector systems on sounding rocket payloads often require interfacing asynchronous outputs with a synchronously clocked telemetry stream. Though this can be handled with an on-board computer, there are several low cost alternatives including custom hardware, microcontrollers, and Field-Programmable Gate Arrays (FPGAs). 
This paper outlines how a telemetry interface for detectors on a sounding rocket with asynchronous parallel digital output can be implemented using low cost FPGAs and minimal custom hardware. 
Low power consumption and high speed FPGAs are available as Commercial Off-The-Shelf (COTS) products and can be used to develop the main component of the telemetry interface. Then, only a small amount of additional hardware is required for signal buffering and level translating. 
This paper also discusses how this system can be tested with a simulated telemetry chain in the small laboratory setting using FPGAs and COTS specialized data acquisition products.

\end{abstract}

\keywords{instrumentation: detectors, instrumentation: miscellaneous, methods: miscellaneous, site testing, space vehicles: instruments}

\section{Introduction}
\label{sect:intro}
Many fields of research within space physics and astronomy utilize experimentation on sounding rockets through the National Aeronautics and Space Administration (NASA). Often, the payload cannot be reliably retrieved and, if sufficient bandwidth is available, data obtained by transducers are telemetered to a ground station via Radio Frequency (RF). In the context of rocket payloads, these transducers could be synchronous monitors for electrical housekeeping or photon counting detectors that make measurements asynchronously. To reduce cost, NASA's Wallops Flight Facility uses a common TeleMetry (TM) system for each sounding rocket. This system offers many standard interfaces, but depending on transducer output format, it may be necessary to implement an electrical TeleMetry InterFace (TMIF). A wide range of methods for this exist, including using custom hardware, microcontrollers or a rugged computer. Another option is the use of Field-Programmable Gate Arrays (FPGAs). FPGAs have the benefit of being highly flexible and re-configurable while also having low power consumption and small Printed Circuit Board (PCB) footprint at a low cost. 

In this paper we describe a TMIF that we have implemented for the Off-plane Grating Rocket for Extended Source Spectroscopy~\cite[OGRESS;][]{Rogers13} using Commercial Off-The-Shelf (COTS), low cost FPGAs and minimal additional hardware. COTS FPGAs serve as the main component of this interface, necessitating custom hardware to be used only for signal buffering and level translating.
The goal of this paper is to provide the reader with a conceptual overview of TM that accords with Inter-Range Instrumentation Group (IRIG) 106\footnote{Refer to www.irig106.org for IRIG 106 documentation.}, and to provide a guideline for designing a TMIF with FPGAs for detectors with asynchronous parallel digital output. First, we outline a Pulse Code Modulation (PCM) formatted~\cite{TMapp06,Plonus01} TM chain, starting with synchronous data collection on board a payload and ending with data reconstruction at a ground station. Next, we describe in detail how we interface our asynchronous detectors with a synchronous PCM formatted TM chain and describe how TMIF was integrated into the OGRESS system. Finally, we demonstrate how this can be simulated in the small laboratory setting using additional FPGAs and COTS products for PCM data acquisition and discuss our conclusions.

\section{Telemetry chain overview}
\label{sect:TM_chain}
The simplest example of a TM chain starts with a single transducer\footnote{In this context, a transducer can be generalized to be any type of device or electrical circuit for spacecraft housekeeping with an analog output that may be sampled synchronously. Examples include, but are not limited to, voltage monitors for power supplies, pressure transducers and thermistor voltage divider circuits that monitor temperature. An example of a transducer that cannot be sampled synchronously in a straightforward manner is an X-ray detector that outputs electrical signals intermittently, every time a photon is detected from an astronomical source. TM interfacing for a transducer with asynchronous output, such as an Gaseous Electron Multiplier (GEM) or a photodiode is discussed in section~\ref{sect:TMIF}} 
on a payload making a measurement of some physical phenomenon and outputting an analog voltage signal. Various techniques exist to transmit these data on an RF transmission link. Common techniques for analog transmission include amplitude modulation and frequency modulation~\cite{Horowitz89}. Analog signals from multiple transducers may be incorporated as different channels on a single transmission link through the process of Frequency Division Multiplexing~\cite[FDM;][]{Plonus01}. Here, each transmitted signal is separated in the frequency domain within a single wide-band link. In practice however, analog modulation techniques pose limitations for a TM system. Not only are analog signals highly susceptible to noise, but FDM also quickly becomes impractical for a system with a large number of channels. Digital modulation techniques have been more commonly used for space applications since the 1970's. In addition to providing noise-resistant transmission, digital modulation techniques offer an efficient way for many channels to be incorporated into a single transmission link.

Digital broadcasting requires data to be `commutated', where analog voltage signals are sampled and digitized by an Analog-to-Digital Converter~\cite[ADC;][]{Horowitz89,Wertz99}. Consider $N$ analog voltage signals, each from a different transducer on a payload, that are to be telemetered to a ground station as different channels. Each signal must be sampled at a rate $S$ that satisfies the Nyquist-Shannon sampling criterion~\cite{Shannon49} and digitized by the ADC to a resolution of $W$ bits to be accurately reconstructed, or `decommutated', at the ground station. FDM may be used to transmit the commutated data from each signal on a different RF carrier. However, sampling events for each signal may also be staggered in the time domain and integrated into a single signal path through the process of Time Division Multiplexing~\cite[TDM;][]{Plonus01,TMapp06}. With TDM, each sampled signal can be transmitted on a single RF carrier as a serial stream of bits. This repeating pattern is partitioned into $N$ slots, also known as words~\cite{Mano08}, that are each $W$ bits wide. Each of these slots contains a digitized sample from one of the $N$ analog signals. This technique is known as Pulse Code Modulation~\cite[PCM;][]{Plonus01,TMapp06}.

\begin{figure}
\centering\includegraphics[scale=0.5]{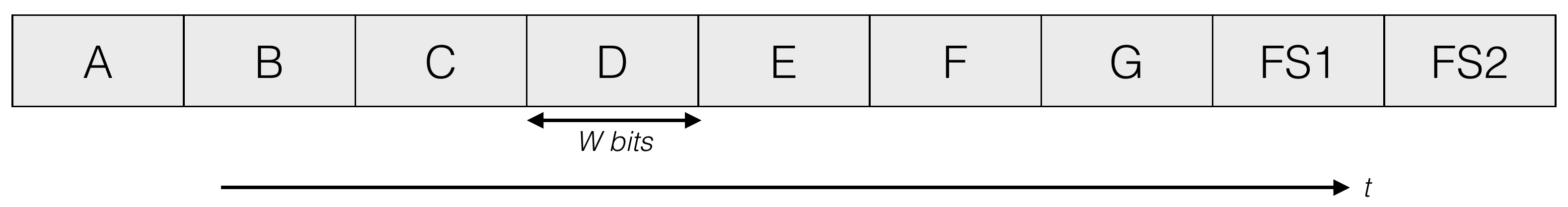}
\caption{Example of a PCM frame, a continously repeating serial bit stream, with $N=9$ slots (words). The 7 slots labeled `A', `B', `C', etc. represent samples from different transducers, each digitized to $W$ bits. Each of these slots occur once per frame, at the normal commutatation rate $S$. The last two slots in a frame are the FS words 1 and 2. Concatenated together, FS1 and FS2 reperesent a unique $2 W$-bit pattern that marks the end of a frame.}\label{fig:FS}
\end{figure}

PCM combines many digitized signal samples into a single continuously repeating pattern of $N$ slots known as a `frame' (see Figure~\ref{fig:FS}). Each of the $N$ slots appear once per frame and the sampling rate $S$ is thus equal to the frame transmission rate. The bit rate $B$ for the PCM stream that follows is given by
\begin{equation}\label{eq:bit_rate}
\frac{B}{\text{bps}}=\frac{N}{\text{slots}}\times\frac{W}{\text{bits/slot}}\times\frac{S}{\text{Hz}} \: .
\end{equation}
Room at the end of a frame is reserved for a unique fixed bit pattern known as the frame synchronization pattern~\cite{Scholtz80} or `Frame Sync' (FS). It is standard for FS to be $2W$ bits wide,\footnote{See IRIG 106 Appendix~C for optimum frame synchronization patterns.} which then replaces two of the $N$ slots in every frame. By detecting FS regularly at a rate $S$, the receiving system achieves `frame lock'. This is required to ensure reliable reception of the PCM stream and to correctly identify each of the $N-2$ remaining slots that contain transducer data. If frame lock is lost, the receiver will lose sampling data for a time period of $\sim S^{-1}$, until the next FS pattern occurs. Though increasing the rate at which the FS pattern occurs improves the reliability of data reception, decreasing this rate leaves more bandwidth for telemetered data. As a result, PCM frame organization must take into account the trade-off between reception reliability and bandwidth usage. 

\begin{figure}
\centering
\includegraphics[scale=0.5]{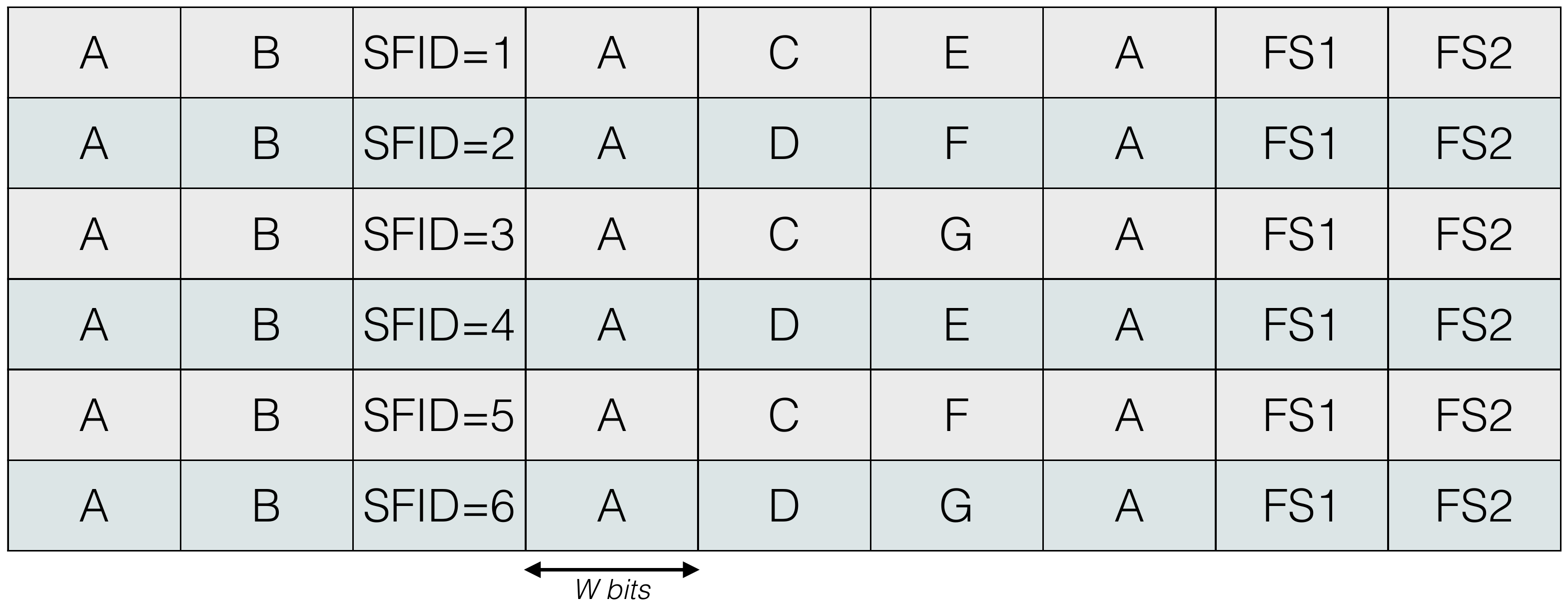}
\includegraphics[scale=0.443]{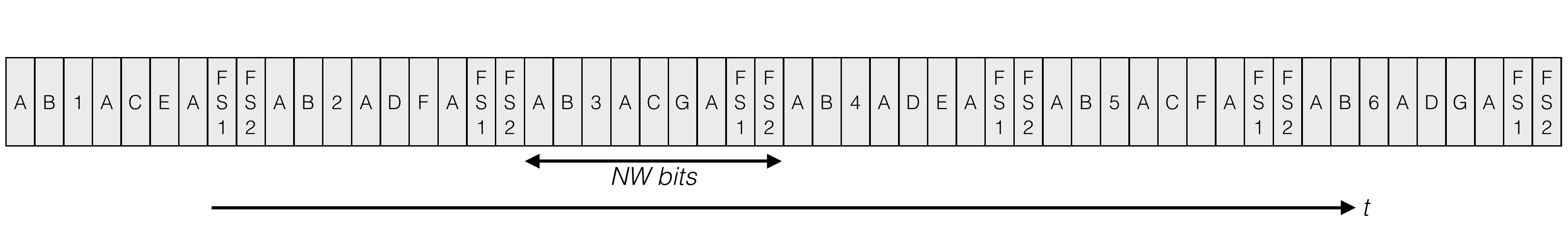}
\caption{Example of a PCM major frame visualized as a matrix (\emph{top}) with $N=9$ columns (words), $M=6$ rows (minor frames/subframes) and 54 total slots. In reality, each of the six $9W$-bit long subframes, defined by the occurrence of FS at a rate $S$, are concatenated together to form a serial $54W$-bit long pattern (\emph{bottom}) that repeats at a rate $S/6$. The third word contains the result of the SFID counter, which distinguishes between subframes within a major frame. The 36 slots labeled `A', `B', `C', etc. represent samples from different transducers, each digitized to $W$ bits. Sample `A' appears three times in every subframe and thus occurs at a super-commutative rate $S^{(3)}_{\text{super}}$. Sample `B' appears once per subrame, and thus occurs at the normal commutative rate $S$. Samples `C' and `D' appear once every two subframes and thus occur at a sub-commutative rate $S^{(2)}_{\text{sub}}$. Likewise, samples `E', `F' and `G' appear once every three subframes and thus occur at a sub-commutative rate $S^{(3)}_{\text{sub}}$.}\label{fig:major_frame}
\end{figure}

A signal being sampled at a rate $S$, occupying the same single slot in each frame, is referred to as `normal commutation'. A signal may also be sampled at a rate $S_{\text{super}}^{(n)}  > S$, given by
\begin{equation}
S_{\text{super}}^{(n)} \equiv nS \: ,
\end{equation}
where $n$ is restricted to be an integer factor of the number of slots $N$. Such a signal thus appears $n$ times in a frame, every $N/n$ slots. This is referred to as `super-commutation'. More sampling flexibility is possible if the organization of the PCM stream is generalized from a single row of $N$ slots to an $M \times N$ matrix known as a `major frame' (see Figure~\ref{fig:major_frame}). In reality, this repeats as a serial pattern of $MN$ slots at the major frame rate $S/M$.
Hereafter, a row of $N$ slots will be referred to as a `minor frame' or `subframe' and an entire column of $M$ slots will be referred to as a word.
Then, assuming the Nyquist-Shannon sampling criterion is still satisfied, a signal may be sampled at a rate $S_{\text{sub}}^{(m)} < S$, given by
\begin{equation}
S_{\text{sub}}^{(m)} \equiv \frac{S}{m} \: ,
\end{equation}
where where $m$ is restricted to be an integer factor of the number of rows $M$. Such a signal thus appears in a single word every $m$ minor frames. This is referred to as `sub-commutation'. Here, the receiver is required to distinguish each of the $M$ minor frames. This is possible with the implementation of a counter which increments with each minor frame and resets with each major frame. The result of this counter, the SubFrame IDentification (SFID), is displayed as a word in the major frame. Together, FS and SFID allow the receiving system to pick up anywhere within a major frame and identify each slot in the matrix correctly.

TM systems for sounding rocket payloads that accord with IRIG 106 currently are provided by NASA's Wallops Flight Facility. The integral component of the system is a PCM encoder, which serves the purpose of commutating and multiplexing data into the electrical PCM stream that is to be transmitted via RF (see Figure~\ref{fig:overall_setup}). A common PCM RF modulation technique for sounding rockets, which we focus on in this paper, is PCM/FM. Based on frequency-shift keying~\cite{Wertz99}, PCM/FM represents binary logic levels using two frequencies $\nu_{\pm} = \nu_0 \pm \Delta \nu$ that deviate from the RF carrier $\nu_0$ by some frequency interval $\Delta \nu$. To reduce wide-band frequency components in the signal, PCM/FM requires phase to be continuous between frequency transitions. This is accomplished by sending the electrical PCM stream to a single voltage-controlled oscillator that varies the signal smoothly between frequency states before transmission~\cite{TM_RF08}. To minimize bandwidth usage further, Non-Return-to-Zero~\cite[NRZ;][]{TMapp06} line coding conventions are often used with PCM/FM. Here, the encoded signal has no rest state at $\nu_0$ and does not change state on timescales shorter than $B^{-1}$. It is known that for NRZ coding, the optimum frequency deviation~\cite{Kotelnikov60,Cartier77,TMapp06} $\Delta \nu_{\text{NRZ}}$ is given by 
\begin{equation}
\frac{\Delta \nu_{\text{NRZ}}}{\text{Hz}} = \frac{\nu_+ - \nu_-}{\text{Hz}} \approx 0.35 \times \frac{B}{\text{bps}} .
\end{equation}
Neglecting any remaining wide-band frequency components, the required bandwidth is $2 \Delta \nu \propto B$.  As a result, RF spectrum allocation must be taken into account when selecting data rates. For example, if FDM is used to integrate multiple PCM streams on different RF carriers into a single transmission link, low bit rates may need to be used to reduce interference.

At the ground station, the receiving antenna system converts PCM/FM  back into an electrical bit stream (see Figure~\ref{fig:overall_setup}). Because NRZ is not inherently self-clocking, the transmitting and receiving systems must agree on $B$ beforehand. Bit edges in the received PCM stream must be synchronized with a ground station clock with a bit rate $B$. A commonly used NRZ coding convention\footnote{See IRIG 106 Chapter~4 for more information on NRZ coding conventions.} is NRZ-Level~\cite[NRZ-L;][]{TMapp06}. Here, physical level transitions follow exactly the logical levels of the PCM stream where a high voltage represents one state while a low voltage represents the other. Correspondingly, in PCM/FM NRZ-L, $\nu_+$ represents one binary state while $\nu_-$ represents the other. A disadvantage of using NRZ-L is that bit synchronization becomes impractical if long strings of the same binary state occur. One way to remedy this issue is to randomize NRZ-L. Known as RNRZ-L, this coding convention utilizes a 15-stage shift register and bit adders~\cite{Horowitz89} to produce a psuedo-random bit sequence based on the input NRZ-L\footnote{See IRIG 106 Appendix~D for a detailed description of this algorithm.}. Broadcasting PCM/FM RNRZ-L greatly increases the number of bit transitions and improves bit synchronization. 

The processes of bit synchronization and frame synchronization can be carried out with specialized PCM acquisition hardware, such as the TarsusHS-PCI$^\text{TM}$ processor board from Ulyssix Technologies\footnote{http://www.ulyssix.com/\#!tarsus-hs-pci-01/cwas}. After bit synchronization, RNRZ-L is decoded to NRZ-L, and the processor checks that the correct FS pattern occurs at the minor frame rate $S$. This hardware coupled with specialized software, usually implemented on a Dewetron$^\text{TM}$ unit\footnote{http://www.dewetron.com/}, is used by Wallops Flight Facility to broadcast data according to the telemetry standard defined in IRIG 106 Chapter~10: each minor frame is encapsulated by a standard header and broadcasted via Ethernet as an IRIG 106 telemetry packet. This provides a way for users without specialized PCM acquisition hardware to decommutate and display data on a personal computer. However, there are benefits of purchasing a TarsusHS-PCI$^\text{TM}$ processor board to use with a personal computer. With knowledge of the PCM organization, the corresponding data acquisition software, DeweSoft$^\text{TM}$\footnote{http://www.dewesoft.com/products/dewesoft-x}, can be easily configured to identify data in the matrix, implement calibrations for digital-to-analog conversion and record the data. The software also contains built-in tools for implementing a Graphical User Interface (GUI) to display the decommutated data (see Figure~\ref{fig:flight_GUI}). The TarsusHS-PCI$^\text{TM}$ processor board and DeweSoft$^\text{TM}$ are also useful for TMIF troubleshooting and TM chain simulation in the small laboratory setting. This is discussed in section~\ref{sect:lab_testing}. 

\section{Telemetry interfacing with parallel digital output}
\label{sect:TMIF}
In practice, not all transducers on a rocket payload will have an analog output. As a result, signal interfacing with the PCM encoder will in general be less straightforward than the scenario described in section~\ref{sect:TM_chain}. The current standard encoder used by Wallops Flight Facility, WFF93\footnote{This encoder was developed by the Telemetry Products and Solutions team at the New Mexico State University Physical Science Laboratory. More information can be obtained from the Wallops Flight Facility Sounding Rockets Program Office; the encoder manual is available upon request. The NASA Sounding Rocket Program Handbook is also available for download at http://sites.wff.nasa.gov/code810/nsroc.html.}, consists of a stack of modular components, or `decks', that support various data formats. The analog deck, which commutates and multiplexes each analog signal into the PCM stream, is just one module of the encoder. Another commonly used module is the parallel deck, which can interface with a transducer that has a parallel digital output. The functionality of the WFF93 parallel deck is standardized to interface directly with synchronous data only, and is not usually readily re-configurable. As a result, for a transducer with an asynchronous parallel digital output, an external TMIF must be implemented to integrate data into the synchronously clocked PCM stream. 

\begin{figure}
\centering
\includegraphics[scale=0.355]{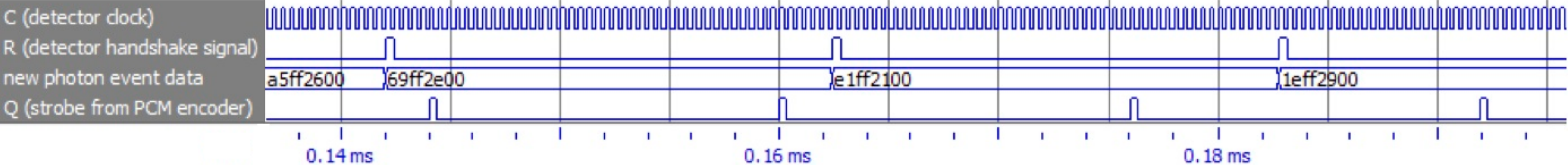}
\caption{Simulated timing diagram showing the signals used for TM interfacing, generated using ModelSim-Altera$^{\text{TM}}$. A clock $C$ running at 2.5~MHz is used for synchronization of the detector electronics. A handshake signal $R$ is asserted by the detector electronics for each new photon event. $R$ is synchronous with the rising edge of $C$ and is asserted for one cycle (0.4~$\mu$s). Here, data for each photon event are represented with $w=32$ bits (expressed in hexadecimal). This output bus changes state whenever a new photon is detected, synchronized with the next closest $C$ rising edge. A strobe signal $Q$ is generated by the PCM encoder. With a $B=10$~Mbps PCM data rate, $Q$ is asserted for a time interval of $3B^{-1}=0.3$~$\mu$s. $Q$ repeats at a rate $S_Q=62.5$~kHz, corresponding to the rate at which photon data can be integrated into the PCM stream in this example.}\label{fig:handshake_signals}
\end{figure}

For the following discussion, consider a photon counting detector that detects incident photons intermittently, where information from each event is represented with $w$ bits. The detector electronics are assumed to have a $w$-bit wide output that changes state for each new photon event. Each state change happens synchronously with the rising edge of an externally provided clock~\cite{Plonus01,Mano08} $C$. Additionally, for each state change, the detector electronics assert a digital handshake signal $R$ for one cycle of $C$. Because it is not assumed that $C$ is synchronized with the encoder in any way, the output of the detector electronics can be taken to be asynchronous relative to the synchronous PCM stream. The encoder needs to integrate the $w$-bit detector data into the PCM stream at some regular rate that corresponds with $S$, $S_{\text{super}}^{(n)}$ or  $S_{\text{sub}}^{(m)}$. The parallel deck of the encoder facilitates data transfer with an auxiliary digital signal known as a `strobe'~\cite[$Q$ hereafter;][]{Mano08}. $Q$ is a short pulse that is asserted for a time interval $3 B^{-1}$ and repeats at a regular rate $S_Q$, corresponding to the rate at which the encoder latches data from the parallel deck port. Figure~\ref{fig:handshake_signals} shows a simulated timing diagram which summarizes these signals, from the detector electronics and the encoder, that are directly involved with the TM interfacing process.

\begin{figure}
\centering
\includegraphics[scale=0.53]{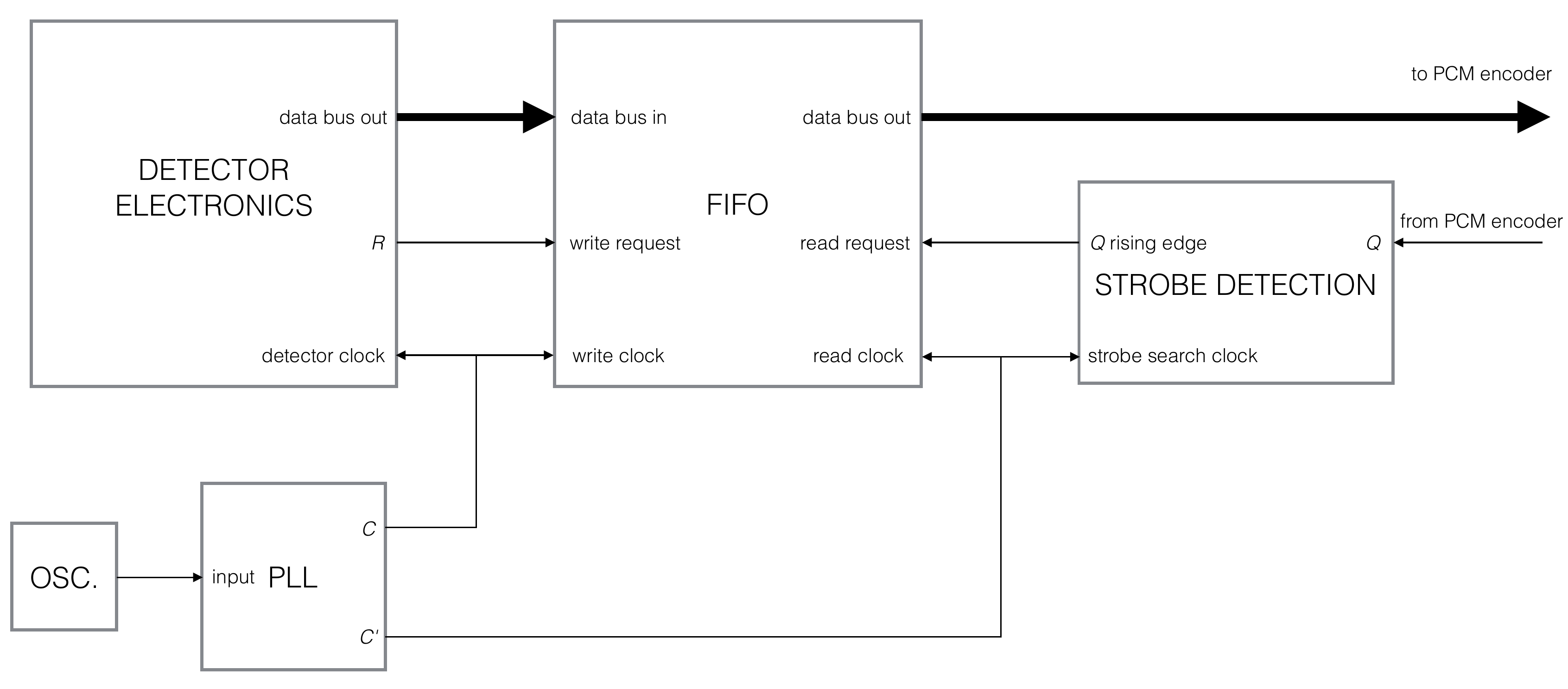}
\caption{Schematic illustrating a basic TMIF design. A PLL with an oscillator input is used to generate clocks $C$ and $C'$. The parallel data bus (represented by the bold arrow) and handshake signal $R$ from the detector electronics are synchronized with $C$, but are asynchronous relative to the strobe $Q$ generated by the PCM encoder. The data are written into a dual-clock FIFO using $R$ as the write request and $C$ as the write clock. A strobe detection module runs off the fast clock $C'$ and outputs a signal $Q_{\text{edge}}$ that is asserted for one cycle of $C'$. The FIFO then uses $Q_{\text{edge}}$ as the read request and $C'$ as the read clock. Parallel data are read from the FIFO at a rate $S_Q$ and are integrated into the synchronously clocked PCM stream by the encoder.}\label{fig:TMIF_schematic}
\end{figure}

The purpose of TMIF is to pass parallel data from the detector to the encoder using the handshake signal $R$ generated by the detector electronics and the strobe signal $Q$ from the encoder. There are many possible ways to design a digital asynchronous-to-synchronous interface for TMIF. Here we describe a design which utilizes a dual-clock, $w$-bit wide First In, First Out~\cite[FIFO;][]{Horowitz89} as the central component (see Figure~\ref{fig:TMIF_schematic}). The FIFO, which acts as a data buffer between the detector electronics and the encoder, requires write clock and write request inputs on the detector side and read clock and read request inputs on the encoder side. The detector side should be synchronized with the detector electronics; both should run off $C$. $R$ can be used as the write request; when $R$ is asserted, the $w$-bit parallel data from the photon event is written to the FIFO. Data from each photon event are stored in the FIFO in sequential order until they are read on the encoder side. 
The encoder provides the strobe $Q$ but does not provide a clock for the read side of the FIFO under normal configuration. As a work-around, a strobe detection module may be implemented where a fast clock $C'$ ($\sim 100~\text{MHz}$) is used to detect the rising edge of $Q$ with high temporal accuracy. Similar to $R$, this detected edge signal $Q_{\text{edge}}$ is asserted for one cycle of $C'$. $Q_{\text{edge}}$ and $C'$ may then be used as the read request and the read clock on the encoder side of the FIFO, respectively. The parallel data are read, one photon event at a time, from the FIFO at a rate $S_Q$ and integrated into the PCM stream by the encoder. However, the expected average rate of $R$ (i.e. the expected photon count rate), $\overline{S}_R$, should be taken into account when designing TMIF. If $\overline{S}_R \leq S_Q$, the FIFO can be designed to be deep enough to prevent data loss. If $\overline{S}_R > S_Q$, the FIFO will overflow regardless of depth, and data will be lost. For the latter case, some other on-board processing system may need to be used.

TMIF is a purely digital system which can be implemented on a variety of logic device platforms. Here we wish to elucidate the benefits of using Field-Programmable Gate Arrays (FPGAs). FPGAs are highly flexible and re-configurable devices that are programmed by the developer with a high level Hardware Description Language~\cite[HDL;][]{Mano08}. The two main languages are Verilog HDL and VHDL, which are used to design a modular, abstract digital system which maps directly to the architecture of the FPGA hardware. Most modern FPGAs manufactured by Xilinx$^\text{TM}$, Inc. and Altera$^\text{TM}$ Corporation are available as development boards, which include a wide range of hardware that is complementary to the FPGA chip itself. The accompanying FPGA design software allows the user to edit HDL, define Input/Output (I/O) routing, simulate the logic design, and re-configure the device in the field. Further, a wide variety of Intellectual Property (IP) from Xilinx$^\text{TM}$ and Altera$^\text{TM}$ exists. This includes blocks of pre-written HDL which greatly reduces design development time. For example, FIFO and Phase-Locked Loop~\cite[PLL;][]{Horowitz89} design functions are both available as flexible and user-friendly soft IP. The TMIF design discussed above can be implemented using the provided FIFO function as the central modular component. The PLL function can take an input oscillator, usually provided on the FPGA development board, to implement clocks $C$ and $C'$. Then, only a small amount of custom HDL is needed to create a strobe detection module and to configure the top level module design. Many low cost FPGAs are on the market, including the Artix$^\text{TM}$\footnote{http://www.xilinx.com/products/silicon-devices/fpga/artix-7.html} and Cyclone$^\text{TM}$\footnote{https://www.altera.com/products/fpga/cyclone-series/cyclone/overview.html} product families from Xilinx$^\text{TM}$ and Altera$^\text{TM}$, respectively. Because they are based in static random access memory, these FPGAs are volatile devices~\cite{Plonus01}. As a result, external flash-based memory is required to re-configure the device upon a power cycle. However, this process is user-friendly with Xilinx$^\text{TM}$ and Altera$^\text{TM}$ design software, and high speed, flash-based storage is standard on most FPGA development boards. 

\section{The OGRESS system}
\label{sect:ogress}
\begin{figure}
\centering
\includegraphics[scale=0.34]{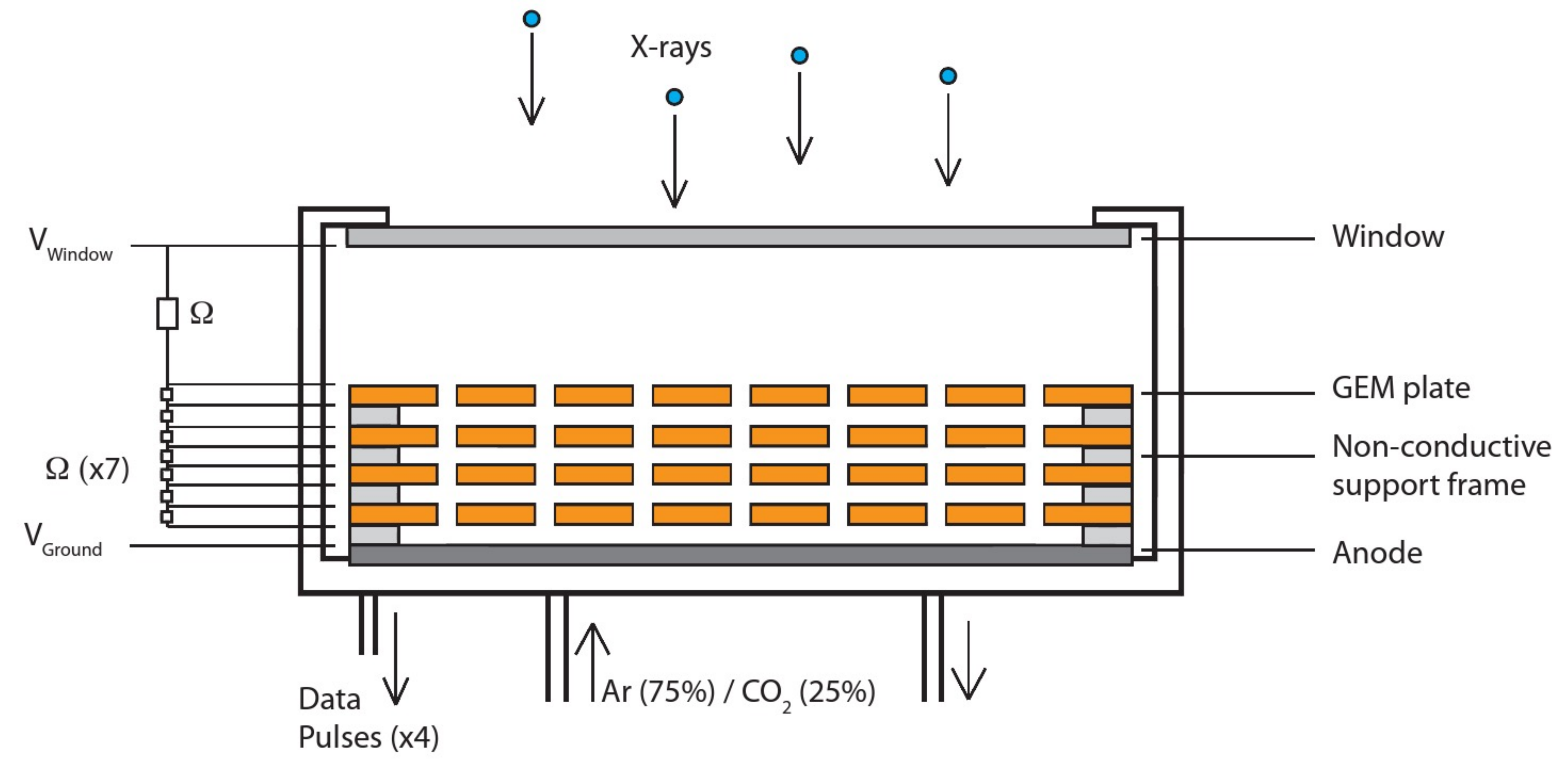}
\caption{Schematic of a GEM detector~\cite{Oakley11}. Collimated X-rays are incident on a 500~nm polyimide window, supported by a stainless steel grid. A 30~nm layer of carbon coating makes the window conductive, which allows it to be held at roughly $-3.5$~kV. The detector is filled with a gas mixture, held at 10~psia, consisting of 75\% Ar, which provides the reservoir for the detection medium, and 25\% CO$_2$, which provides charge replenishement. An incident X-ray photoionizes the Ar gas and the produced charge is accelerated toward a grounded anode, where the charge signal is detected. In between the window and the anode are a series of porous GEM plates held at increasing voltages using a resistor chain. As charge is acceleared through the pores of GEM plates, a cascade of electrons is produced. The resulting electron shower is an amplified version of the initial signal. The anode consists of two independent cross-delay lines that send charge pulses to a TDC that infers the $x$ and $y$ coordinates and the charge pulse height of each photon event.}\label{fig:GEM}
\end{figure}

The Off-plane Grating Rocket for Extended Source Spectroscopy~\cite[OGRESS, NASA sounding rocket 36.292;][]{Rogers13}  was successfully launched from White Sands Missile Range in May 2015. OGRESS is the fourth generation of a wide-field soft X-ray spectrograph coming from the Center for Astrophysics and Space Astronomy at the University of Colorado, Boulder~\cite{McEntaffer08,Oakley09,Zeiger11}. To make a spectral observation of the Cygnus Loop supernova remnant, the payload utilized wire-grid collimator optics~\cite{Shipley11}, off-plane diffraction grating arrays~\cite{Cash1991,McEntaffer07} and Gaseous Electron Multiplier (GEM) detectors~\cite{Zeiger13} manufactured by Sensor Sciences, LLC\footnote{http://sensorsciences.com/}. GEM detectors are large format, gas-filled detectors which detect X-rays on a photon-by-photon basis by measuring charge produced from photoionization events (see Figure~\ref{fig:GEM}). An X-ray incident on a GEM detector passes through a thin, conductive window held at roughly $-3.5$~kV and creates a photoelectron from the contained argon gas. A series of porous GEM plates~\cite{Sauli97} held at high voltages are placed between the window and a grounded anode, which consists of two independent cross-delay lines. As the photoionized charge is accelerated toward the anode, collisions occur within the GEM plate pores. This causes a cascade of electrons which provides gain on the order of $10^4$ for the initial signal. A Timing-to-Digital Converter (TDC), also manufactured by Sensor Sciences, infers $x$ and $y$ coordinates of the incident photon on the detector by comparing the arrival times from the ends of each anode line. Additionally, total charge from the photon event is measured. The $x$ and $y$ coordinates are both digitized to a resolution of 12 bits and the charge pulse height is digitized to a resolution of 8 bits. These data, concatenated together as a 32-bit parallel data bus, are written to a FIFO internal to the TDC. When a rising edge of the detector clock $C$ is detected, the data are read from the FIFO and routed to the output of the TDC. For each new photon event, a handshake signal $R$ is asserted for one cycle of $C$ and the output bus changes state after a small time delay ($\sim 10$~ns). 

\begin{figure}
\centering
\includegraphics[scale=0.2]{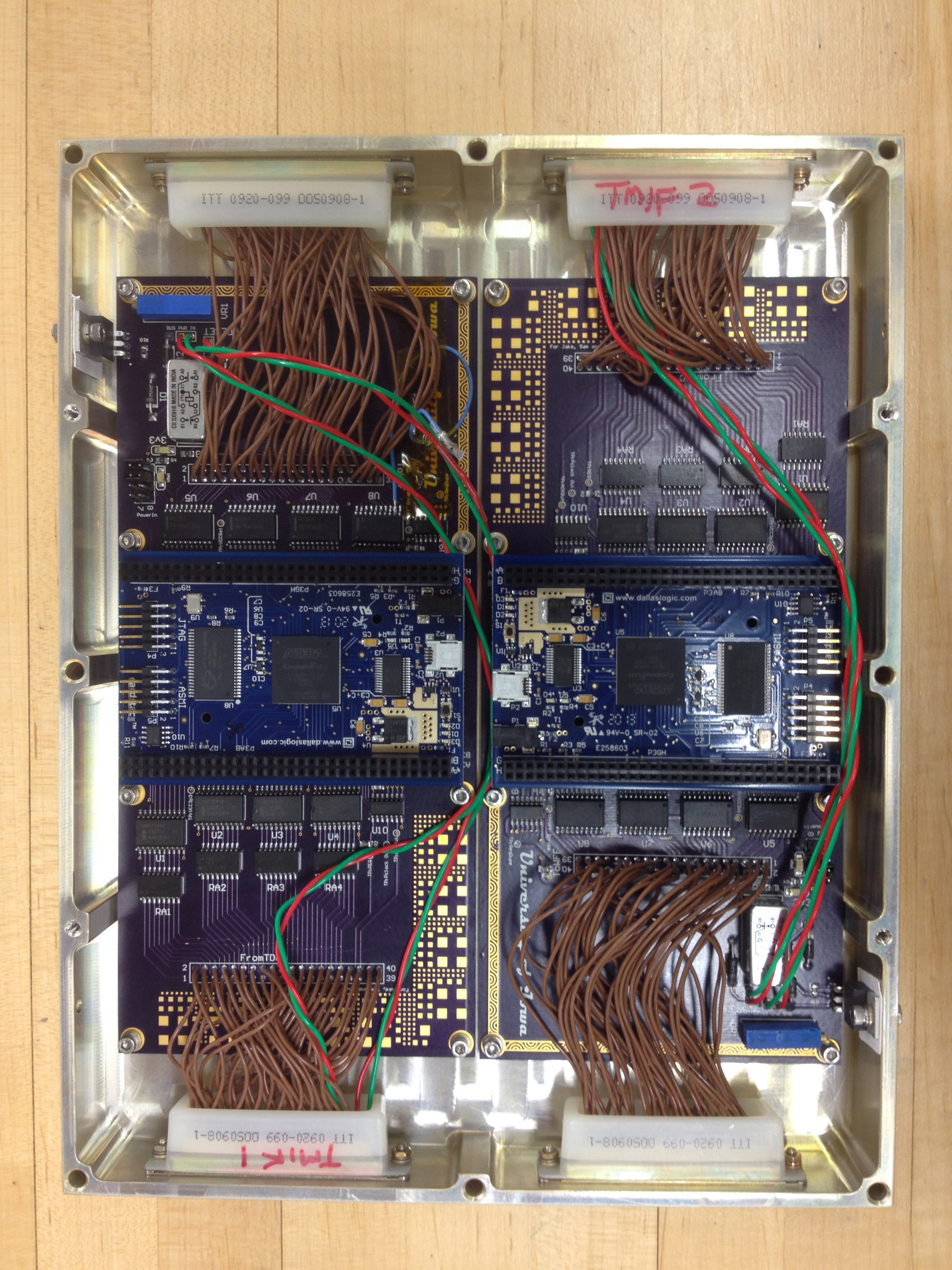}
\caption{Flight hardware for TMIF. Each of the two GEM detectors on the OGRESS payload uses a separate FPGA-based TMIF module. For each channel, a Dallas Logic$^\text{TM}$ CMCS002 module loaded with the TMIF design discussed in section~\ref{sect:ogress} is socketed into custom circuitry consisting of signal buffering and level translating components. The circuitry level translates $C$ (generated by the FPGA) from a 3.3~V to 5~V logic level before it is routed to the TDC.
$R$ and each bit of the TDC data bus are buffered from 5~V to 3.3~V logic levels before they are input to the FPGA. Further, $Q$ is buffered from a 5~V to 3.3~V logic level before it is input into the FPGA and each bit of the synchronous output data bus are level translated from 3.3~V to 5~V before they are routed to the PCM encoder.}\label{fig:TMIF_flight}
\end{figure}

\begin{figure}
\centering
\includegraphics[scale=0.53]{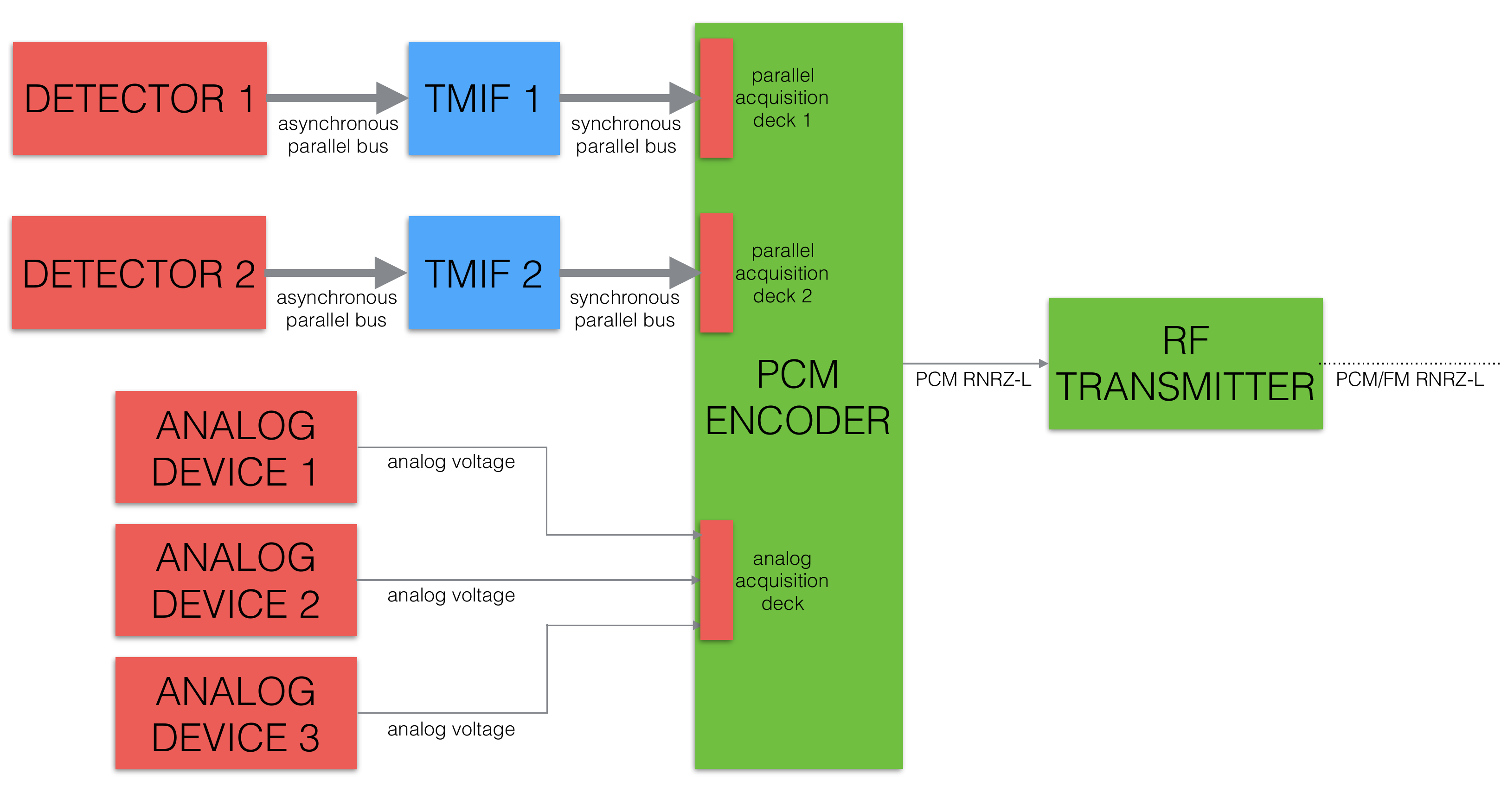}
\includegraphics[scale=0.53]{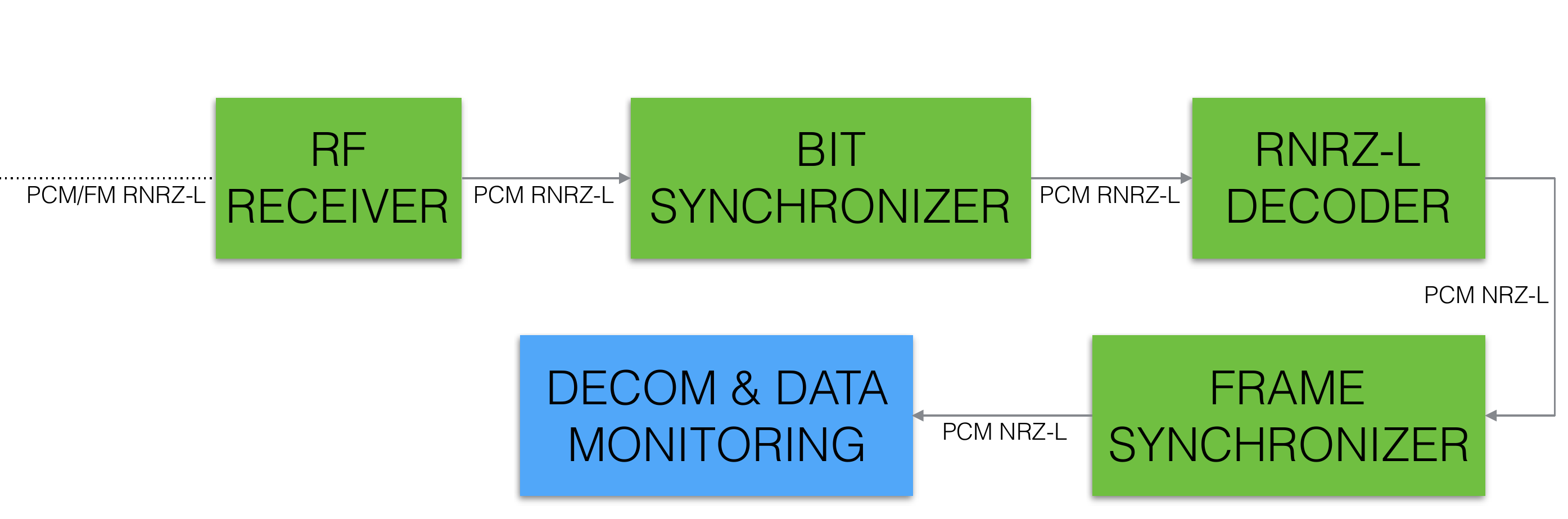}
\caption{Schematic illustrating the data acquisition and TM system on the OGRESS payload (\emph{top}) and at the ground station (\emph{bottom}). Red boxes represent transducers on board the payload. Green boxes represent components provided by Wallops Flight Faciltity or White Sands Missile Range. Blue boxes represent COTS components implemented for TMIF and data monitoring. Each detector outputs asynchronous parallel digital data to an FPGA-based TMIF unit. Each TMIF unit outputs synchronous parallel digital data to a parallel deck of the PCM encoder. 31 transducers for payload housekeeping (for brevity, only three are shown here) output analog signals that are routed to the analog deck of the PCM encoder. There, each signal is digitized by a 10-bit ADC. The PCM encoder integrates data from the parallel and analog decks into a pseudo-randomized PCM bit stream which is transmitted to the ground station via RF. The signal is recieved and converted back into an electrical bit stream at the ground station. The bit rate is synchronized with a ground station clock and the randomized bit stream is decoded to NRZ-L. After frame synchronization, the data are decommutated and displayed (see Figure~\ref{fig:flight_GUI}). }\label{fig:overall_setup}
\end{figure}

To integrate the data into the PCM stream, OGRESS used an FPGA-based TMIF for each GEM detector TDC output (see Figure~\ref{fig:TMIF_flight}). Asynchronous data from each TDC are input into separate TMIF units (see Figure~\ref{fig:overall_setup}). Each TMIF unit outputs synchronous data to a parallel deck of the PCM encoder.
Each TMIF unit, designed and built at the University of Iowa, uses a Dallas Logic$^\text{TM}$ CMCS002 module\footnote{http://dallaslogic.com/cmcs002-controllerfpga-module} which features an Altera Cyclone~III$^\text{TM}$ device. The module also includes an EPCS16 serial flash device for FPGA power-on configuration, a 25~MHz clock oscillator, and I/O headers large enough to fit the 32-bit output bus, $R$ and $C$. While the CMCS002 module works with 3.3~V logic levels, both the TDC and the PCM encoder work with 5~V logic levels. To remedy this, each CMCS002 module was socketed into a custom PCB that consists of signal buffers and level translators. As part of the compilation process, the Altera Quartus~II$^\text{TM}$ software package\footnote{https://www.altera.com/products/design-software/fpga-design/quartus-ii/overview.html} synthesizes the HDL to a gate level description. Because the Cyclone~III$^\text{TM}$ device is volatile, using Quartus~II$^\text{TM}$ the EPCS16 serial flash chip was configured with the gate level description through the Active Serial Memory Interface (ASMI) port of the CMCS002 module using a USB-Blaster$^\text{TM}$ download cable. Then, upon device power-up, the design is automatically loaded onto the Cyclone III$^\text{TM}$ chip, at which point the TMIF unit is fully functional. 

The FPGAs were configured with a design based on that discussed in section~\ref{sect:TMIF}. Using Quartus~II$^\text{TM}$, we developed Verilog HDL that makes use of the FIFO and PLL Altera Megafunctions$^\text{TM}$. The FIFO Megafunction$^\text{TM}$, which was configured to be 32 bits wide and 4,096 words deep, serves as the central component to the design. The PLL Megafunction$^\text{TM}$, using the 25~MHz clock oscillator as an input, generates $C$ at 2.5~MHz and $C'$ at 100~MHz. In addition to being routed internally to the write clock input of the FIFO, $C$ is level translated to 5~V and routed externally to the TDC. $C'$ is routed internally to the read clock input of FIFO and the input of the strobe detection module, which provides the signal $Q_{\text{edge}}$. To account for the small delay between $R$ and the TDC output state change, a signal identical to $R$, but delayed by a half cycle of $C$, is used as the FIFO write request. $Q_{\text{edge}}$ is used for the FIFO read request under the condition that the FIFO is not empty (the FIFO Megafunction$^\text{TM}$ provides this flag). Data read from the FIFO are then stored in a register used for the output of TMIF. The encoder latches these data each time $Q$ is asserted and integrates them into the PCM stream.  If $Q_{\text{edge}}$ is asserted while the FIFO is empty, the data from the previous photon event will be latched by the encoder again, resulting in a duplication. To prevent this, the output TMIF register is masked with a 32-bit string of zeros whenever $Q_{\text{edge}}$ is asserted while the FIFO is empty. A timing diagram generated from a simulation of this protocol is shown in Figure~\ref{fig:TMIF_modelsim}.

The TM system provided by Wallops Flight Facility uses transmission links on two different RF carriers. Only one of these is used for standard PCM/FM; the other is used as a video feed for the celestial attitude control system. The PCM/FM link is broadcast at $B = 8$~Mbps and encoded according to RNRZ-L. The PCM stream has an $M \times N = 32 \times 120$ major frame structure with $W=16$ bits/slot. From equation~\ref{eq:bit_rate}, the minor frame rate is $S = B/NW = 4.166$~kHz and the major frame rate is $S/M = 130$~Hz. The WFF93 PCM encoder provides two strobes, offset in time from one another, for each of the two TMIF units. Each strobe is asserted for a time interval of $3B^{-1}=0.375$~$\mu$s and repeats at rate of $S_Q = 50$~kHz, corresponding to a super-commutative sampling rate of $S_Q = 12 S = S_{\text{super}}^{(12)}$. Because each set of photon event data is 32 bits wide, two words in the major frame are used to encapsulate each sample. If the major frame is visualized as a matrix, data from a single detector occupy two adjacent columns, every 10 columns. The OGRESS payload also has 31 transducers with analog output used for housekeeping of the electronics assembly (see Figure~\ref{fig:overall_setup}). Each of these signals are digitized and sampled at a rate $ S_{\text{sub}}^{(32)}$ and digitized with a 10-bit ADC, contained in the analog deck of the PCM encoder. These samples thus appear as a subword once every major frame. In addition to the analog and parallel decks, the encoder also uses a counter deck which uses $R$ from each detector chain and provides an effective photon count rate averaged over one second. These data are 18 bits wide and occur at a rate $ S_{\text{sub}}^{(8)}$, thereby occupying space over two words every 8 subframes. 

\begin{figure}
\centering
\includegraphics[scale=0.42]{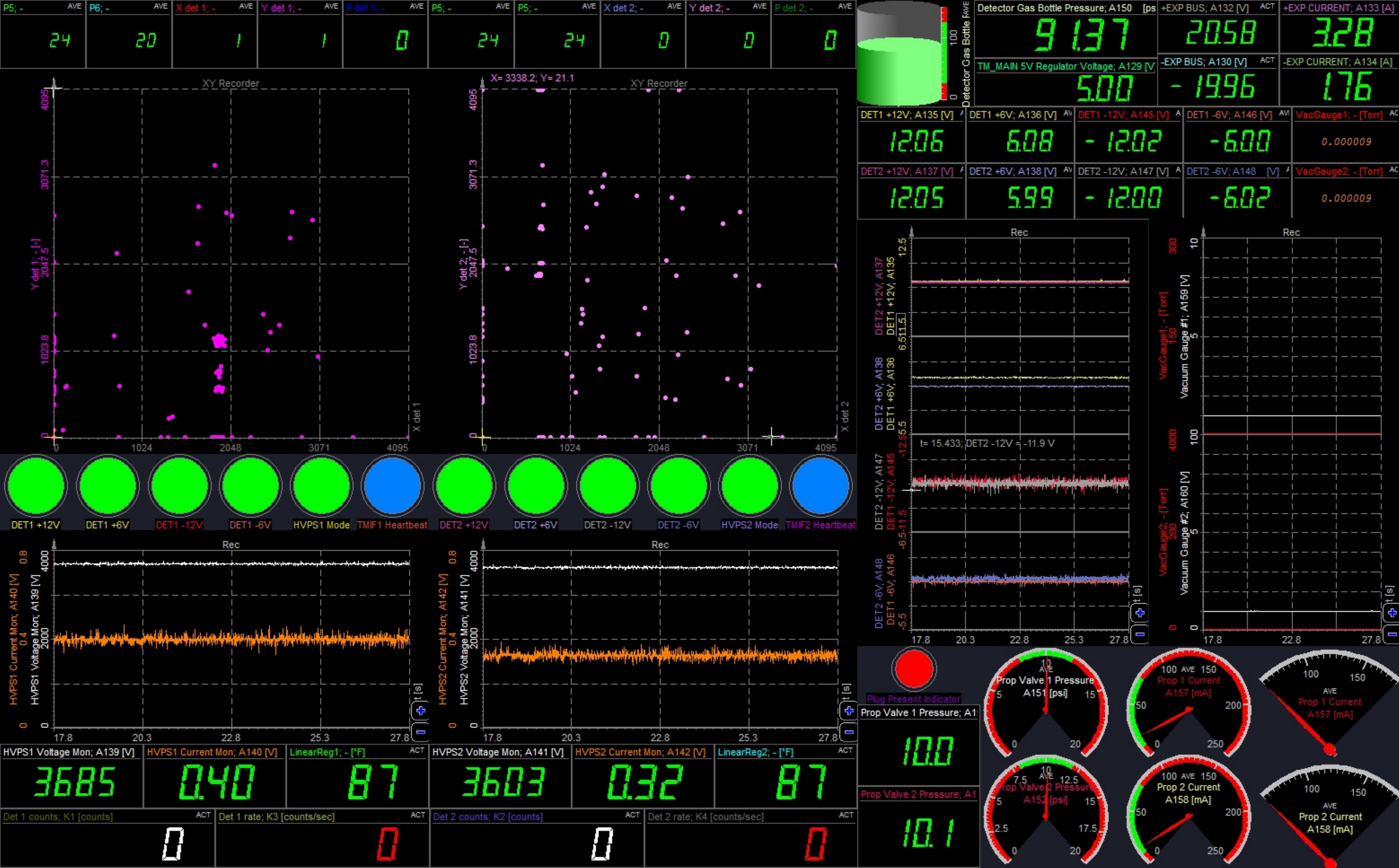}
\caption{Flight data monitoring GUI developed with DeweSoft$^\text{TM}$. The top left portion displays data from each GEM detector. The top left row of boxes show these data as continuously updating numerical values while
extracted $x$ and $y$ data are plotted cumulatively on two-dimensional scatter plots. Below, lamp icons indicate whether low volt supplies for each detector are nominal (green) or out of range (red). Additionally, a blue lamp icon flashes with the $\sim 60$~Hz heartbeat signal generated by each TMIF module. The bottom left portion shows detector high volts data in the form of recorder plots and as continuously updating numerical values. Below, data from the counter deck of the encoder are displayed. The top right portion shows data for the detector gas bottle pressure, bus power supplies, detector low volts supplies and payload pressure. The bottom right portion shows the GEM detector pressures and proportional valve currents.
}\label{fig:flight_GUI}
\end{figure}

Computers loaded with DeweSoft$^\text{TM}$ and equipped with TarsusHS-PCI$^\text{TM}$ processor boards were used to monitor data during the flight of OGRESS. These processor boards directly take the received RNRZ-L bit stream as an input. Using DeweSoft$^\text{TM}$, we developed a GUI that plots $x-y$ data for each detector, plots voltages versus time and displays many other channels as dials and monitors (see Figure~\ref{fig:flight_GUI}). In addition, we used custom software developed in the C programming language that takes IRIG 106 telemetry packets as an input, extracts the data and plots them. This software provided scaled $x-y$ plots that binned data and incorporated pulse height information. It also calculated photon count rates based on the occurrence of data in the PCM stream, which provided a means to verify that $R$ was interfacing with the counter deck of the encoder correctly.

\section{Telemetry chain simulation and laboratory testing}
\label{sect:lab_testing}
In the early stages of development, the HDL for TMIF was simulated using the software ModelSim-Altera$^{\text{TM}}$\footnote{https://www.altera.com/products/design-software/model---simulation/modelsim-altera-software.html} in conjunction with Altera Quartus~II$^\text{TM}$, which allowed us to troubleshoot the design and to make iterative improvements to the code. Quartus~II$^\text{TM}$ synthesizes the HDL into a gate level description and ModelSim-Altera$^{\text{TM}}$ simulates the logical function of the design and generates a timing diagram based on a user-provided HDL test bench. For TMIF testing, we developed an HDL module that, using an input clock $C$, generates simulated asynchronous detector data and a handshake signal $R$. In reality, the detector electronics (i.e. the TDC discussed in section~\ref{sect:ogress}) change output state intermittently. However, data generated at a regular rate can be taken to be asynchronous as long as the timing is not synchronized with $Q$ from the PCM encoder. The detector simulator module generates fake photon event data by first using $C$ to increment a counter. When the state of the counter reaches a certain set value, the module asserts $R$ for one cycle of $C$, at which point the counter resets. The rate at which $R$ is asserted, $S_R$, essentially represents $\overline{S}_R$, the expected average photon count rate from the detector. Then, the rising edge of $R$ is used to drive state changes of additional counters that represent the two-dimensional position and pulse height of a photon event. Two 12-bit registers represent the $x$ and $y$ coordinates of a photon event, while an 8-bit register represents pulse height. For simplicity, the two 12-bit registers were designed to increment in an identical fashion so that accumulated data would appear as a line on an $x-y$ plot. Along with $R$, the three registers, concatenated together to form a single 32-bit register, were used to emulate the output of a GEM detector TDC. A timing diagram of this design is shown in Figure~\ref{fig:handshake_signals}.

\begin{figure}
\centering
\includegraphics[scale=0.355]{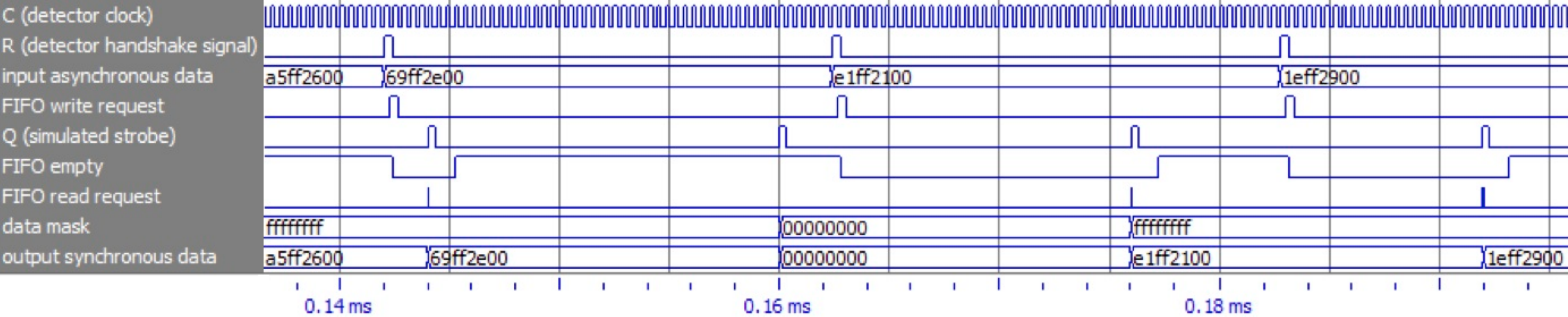}
\caption{Timing diagram depicting the simulated TMIF protocol used for OGRESS, generated using ModelSim-Altera$^{\text{TM}}$. Clock $C$ drives the detector simulator module which generates $R$ for one cycle, and repeats at a rate $\overline{S}_R=S_R=50$~kHz. $R$ is used to increment counters that represent the two-dimensional position and pulse height of a photon event. These data are concatenated together as a 32-bit wide output, which simulates the asynchronous output of a GEM detector. These data are written to the FIFO using a write request signal that is identical to $R$, but delayed by a half clock cycle of $C$. A simulated strobe $Q$ is asserted at a rate $S_Q=62.5$~kHz. $Q_{\text{edge}}$ is used as the read request only when the FIFO is not empty. Data read from the FIFO are stored in a register used for the synchronous output of TMIF. To prevent duplicate data from being latched by the PCM encoder, this register is reset using a 32-bit wide mask if $Q$ is asserted while the FIFO is empty.}\label{fig:TMIF_modelsim}
\end{figure}

To test the design of TMIF with ModelSim-Altera$^{\text{TM}}$, a simulated strobe generator is needed. This was implemented as a basic counter-based HDL module that asserts $Q$ for three cycles of a clock running at $B$ and repeats the signal at a frequency $S_Q$. Then, the detector simulator, TMIF and strobe generator modules were combined into a single top level module and synthesized to a gate level description using Quartus~II$^\text{TM}$. An HDL test bench with input registers for clocks $C$, $C'$ and $B$ was used to generate the timing diagram seen in Figure~\ref{fig:TMIF_modelsim}. Here, $C$ drives the detector simulator module which provides $R$ and the input asynchronous data. The FIFO write request request signal is identical to $R$, but delayed by a half cycle of $C$. The strobe detection module uses $C'$ to provide $Q_{\text{edge}}$, which is used for the FIFO read request if the FIFO is not empty. Data read from the FIFO are stored in register used for the output of TMIF. Further, if $Q_{\text{edge}}$ occurs while the FIFO is empty, the output of TMIF is masked to zero. As discussed in section~\ref{sect:ogress}, this prevents duplicate data from being integrated into the PCM stream. With ModelSim-Altera$^{\text{TM}}$ it was verified that the design of TMIF allows data input at a rate $\overline{S}_R=S_R$ to be output at a rate $S_Q$, in sequential order.

\begin{figure}
\centering
\includegraphics[scale=0.5]{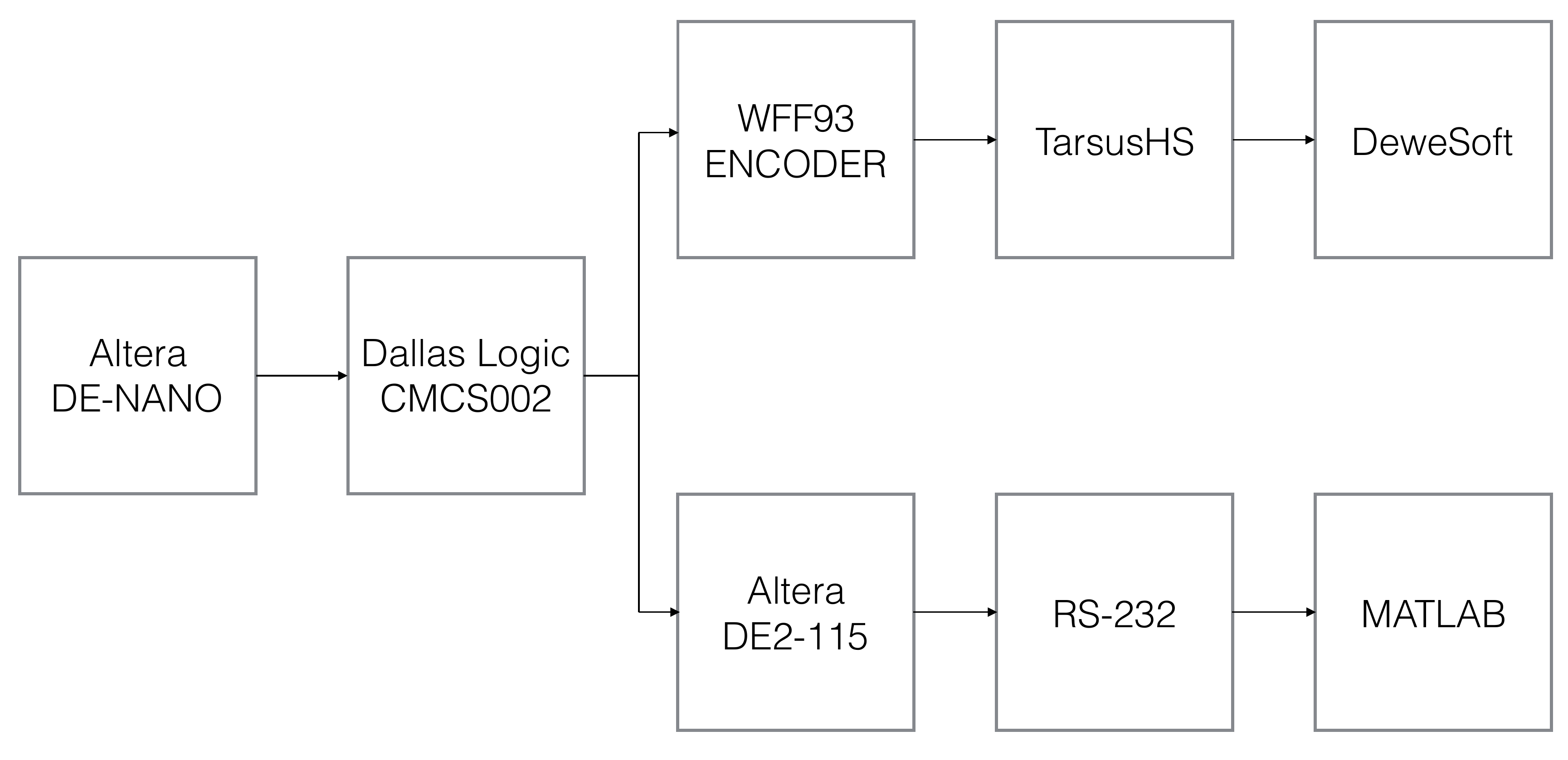}
\caption{Schematic illustrating the overall architecture of the laboratory TM chain simulation. A Cyclone~IV$^\text{TM}$ device on an Altera DE0-Nano Development Board$^\text{TM}$ is configured with HDL to simulate asynchronous detector data. This output is routed to a TMIF unit, which makes use of a Cyclone~III$^\text{TM}$ device on board a Dallas Logic$^\text{TM}$ CMCS002 module. From here, two options exist to simulate the remainder of the TM chain. The first option (\emph{top}) utilizes a WFF93 encoder and specialized COTS products for PCM data acquisition including a TarsusHS-PCI$^\text{TM}$ processor board from Ulyssix Technologies and DeweSoft$^\text{TM}$. This setup, used for both laboratory testing and integration at Wallops Flight Facility, is discussed more thoroughly in Figure~\ref{fig:TM_lab_sim}. The second option (\emph{bottom}) utlizes a Cyclone~IV$^\text{TM}$ device on an Altera DE2-115 Development and Education Board$^\text{TM}$ configured with HDL for a strobe simulator and a parallel-to-serial module. Data are output to a personal computer via RS-232, where they are retrieved and plotted using MATLAB$^\text{TM}$.}\label{fig:test_setup}
\end{figure}

While HDL can be tested extensively with simulation software, it is crucial to verify the functionality of TMIF on a hardware level (see Figure~\ref{fig:test_setup}). To do this, a hardware detector simulator which emulates the 5~V logic level output of a GEM detector TDC was implemented to generate a known signal through a TMIF unit. This was done using an Altera DE0-Nano Development Board$^\text{TM}$\footnote{https://www.altera.com/b/de0-nano-dev-board.html} and a custom PCB consisting of level translators. The Cyclone~IV$^\text{TM}$ device on the DE0-Nano$^\text{TM}$ board, which works with 3.3~V logic levels, was configured with the detector simulator HDL discussed above. 
The FPGA takes $C$ as an input and outputs the 32-bit simulated asynchronous data and $R$ with a 40-pin I/O header on the DE0-Nano$^\text{TM}$ board. Socketing the level translator PCB onto this header translated $C$ from 5~V to 3.3~V and the 32-bit output data and $R$ from 3.3~V to 5~V. Sensor Sciences provides data acquisition hardware and software, DSTAcq$^\text{TM}$, that directly interfaces with the parallel output of a GEM detector TDC. This hardware provides $C$ at 5~V, latches the 32-bit asynchronous data and plots them. By interfacing directly with DSTAcq$^\text{TM}$ and checking that data appeared as expected, it was verified that the hardware detector simulator properly emulates a GEM detector TDC. 

Once its functionality had been verified, the hardware detector simulator was used to pass predictable data through a test TMIF module consisting of a Dallas Logic$^\text{TM}$ CMCS002 module socketed into the custom circuitry discussed in section~\ref{sect:ogress}. One way to interface with TMIF on the encoder side is to implement a module on a third FPGA that serves a strobe generator and a parallel-to-serial transmitter which outputs data to a personal computer. This was done by using an Altera DE2-115 Development and Education Board$^\text{TM}$\footnote{http://www.terasic.com.tw/cgi-bin/page/archive.pl?Language=English\&No=502}, which features a Cyclone~IV$^\text{TM}$ device and many hardware interfaces, including an RS-232~\cite{Horowitz89,Plonus01} port. Because the device works with 3.3~V logic levels, custom circuitry for I/O level translation similar to the hardware detector simulator was required for interfacing. The strobe generator module discussed above was implemented with a clock running at $B$ provided by a PLL Megafunction$^\text{TM}$ that uses a 50~MHz oscillator on the DE2-115$^\text{TM}$ board as an input. The parallel-to-serial transmitter module was implemented using a FIFO Megafunction$^\text{TM}$, open source HDL modules for a baud tick generator and an asynchronous transmitter\footnote{See http://www.fpga4fun.com/SerialInterface.html which provides a tutorial on how to implement an RS-232 transmitter with an FPGA. The HDL provided therein was used as a baseline for this parallel-to-serial transmitter module.}, and minimal custom HDL. The FIFO, configured to have a 32-bit wide input and an 8-bit wide output, serves the purpose of taking the data from TMIF and outputting them to the asynchronous transmitter module, one byte at a time. The asynchronous transmitter module encapsulates each byte with standard start, parity and stop bits. This serial output is then routed to the RS-232 transceiver chip on the DE2-115$^\text{TM}$ board. By broadcasting RS-232 to a personal computer and using MATLAB$^\text{TM}$ to retrieve, plot and examine the data, the functionality of TMIF could be verified.

\begin{figure}
\centering
\includegraphics[scale=0.19]{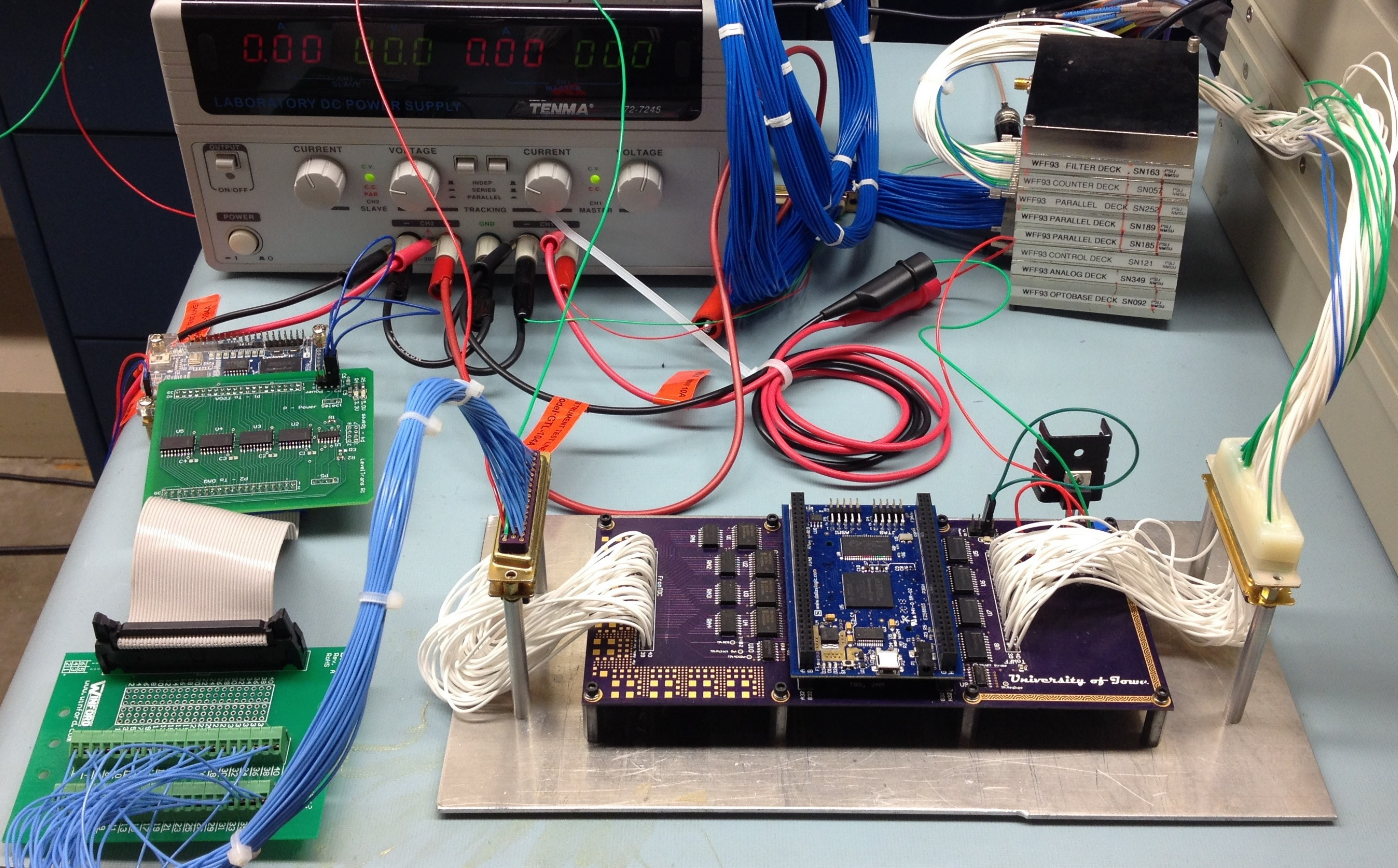}
\caption{TM chain simulation laboratory setup. An Altera DE0-Nano Development Board$^\text{TM}$ with socketed custom circuitry simulates the output of a GEM detector TDC (\emph{left}). The data are passed to a TMIF test unit through a breakout board. A WFF93 encoder, borrowed from Wallops Flight Facility (\emph{right}), reads the data from the TMIF unit and integrates them into the PCM stream. This bit stream is output though a BNC port and routed to the input of a TarsusHS-PCI$^\text{TM}$ processor board installed on a laboratory computer. Finally, the data are decommutated and displayed with DeweSoft$^\text{TM}$.}\label{fig:TM_lab_sim}
\end{figure}

The remainder of TM chain, save RF transmission, was also simulated in the laboratory using a WFF93 encoder obtained from Wallops Flight Facility, a purchased TarsusHS-PCI$^\text{TM}$ processor board and DeweSoft$^\text{TM}$. The WFF93 encoder, configured with the OGRESS PCM matrix, included multiple parallel decks and an analog deck for the two GEM detectors and the analog transducers on the OGRESS payload, respectively. Interfacing with an encoder provided a direct way to verify that data were being integrated into the PCM stream correctly. The WFF93 outputs the RNRZ-L encoded PCM bit stream through a BNC port. This signal was routed directly to the input dongle of a TarsusHS-PCI$^\text{TM}$ processor board where it was bit synchronized, decoded to NRZ-L and frame synchronized. DeweSoft$^\text{TM}$, configured with the OGRESS PCM organization, was used to decommutate and display the data using the GUI discussed in section~\ref{sect:ogress} (see Figure~\ref{fig:flight_GUI}). To verify that TMIF passes data through the simulated TM chain correctly, the TMIF test module, used with the hardware detector simulator, was routed to a parallel deck of the WFF93. Checking that data appeared as expected on a $x-y$ plot provided some assurance that data were being passed correctly. However, verification was taken a step further by using DeweSoft$^\text{TM}$ to visualize the received PCM stream as an $M \times N = 32 \times 120$ matrix, where each cell is a 16-bit number expressed in hexadecimal. In this view, the matrix was manually inspected to ensure that data were received as expected on a count-by-count basis. To provide easy pattern recognition, the detector simulator was re-configured to output data that incremented in powers of two. During this examination process it was discovered that data were not masked by TMIF properly whenever a read request was asserted during a write request. After the HDL was edited to add in this condition for the assertion of the read request signal, it was confirmed that data were being integrated into the PCM stream in sequential order with no duplication. After the functionality of TMIF had been verified through this process, the simulated TM chain was used to acquire data from the GEM detector TMIF units and the analog transducers on the OGRESS payload. This provided a means to calculate digital-to-analog conversions and develop the DeweSoft$^\text{TM}$ GUI that was used for monitoring during flight. Additionally, this laboratory setup was used for post-flight calibrations and testing.

\section{Conclusions}
\label{sect:conclusions}
The TMIF implemented for OGRESS proved to be robust throughout laboratory testing, integration at Wallops Flight Facility and launch at White Sands Missile Range. If the expected average count rate $\overline{S}_R$ is not larger than the strobe rate $S_Q$ from the PCM encoder, this design should be easily transferable to other photon counting detector systems that use parallel digital output. FPGAs, being highly flexible and reliable, provided an efficient and effective platform for the TMIF system. Low power consumption and high speed devices are attainable as COTS products and minimal additional hardware is required to implement a TMIF module suitable for flight. Though vendor-provided soft IP reduces the amount of custom HDL needed to implement the design, development time still hinges on learning on HDL syntax and gaining familiarity with the design and simulation software. However, designing and configuring FPGAs for TMIF aids in the understanding of how data are stored when received at the ground station and how they should look before they are processed for analysis.

FPGAs also provided a convenient, low cost avenue for testing the functionality of TMIF in the small laboratory setting. 
A step further was taken by making use of a PCM encoder borrowed from Wallops Flight Facility and COTS specialized hardware and software for PCM data acquisition. This hardware and software allowed us to complete all handshake testing in-house at the University of Iowa. Furthermore, we saved time and effort during integration at Wallops Flight Facility by using the COTS software to develop the GUI for data monitoring and to perform all digital-to-analog conversions in advance. 
The FPGA design and the PCM hardware will also be used for future sounding rocket projects at the University of Iowa that make use of other photon counting detector systems.

\section*{Acknowledgments}
This work was supported by NASA grant NNX13AD03G. Special thanks are due to Dr. Richard Harriss for the early development of the Verilog HDL code used for TMIF, and Mr. William Robinson at the University of Iowa for the development of the custom data monitoring software used during the flight of OGRESS. We would also like to thank the personnel at Wallops Flight Facility and White Sands Missile Range for their aid in our understanding of telemetry.

\bibliographystyle{ws-jai}

\bibliography{report}

\end{document}